\documentclass[conference]{IEEEtran}
\IEEEoverridecommandlockouts
\usepackage{cite}
\usepackage{amsmath,amssymb,amsfonts}
\usepackage{algorithmic}
\usepackage{graphicx}
\usepackage{textcomp}
\usepackage{xcolor}
\usepackage{academicons}
\usepackage{hyperref}
\usepackage{fancyhdr}

\newcommand{\orcidicon}{\includegraphics[width=0.32cm]{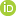}}
\newcommand{\orcid}[1]{\href{https://orcid.org/#1}{\orcidicon}}

\def\BibTeX{{\rm B\kern-.05em{\sc i\kern-.025em b}\kern-.08em
    T\kern-.1667em\lower.7ex\hbox{E}\kern-.125emX}}

\makeatletter
\newcommand{\rightconfheader}{\parbox{\textwidth}{%
    \flushright  
    \rmfamily\fontsize{8}{10}\selectfont
    Proc. of the 5th International Conference on Electrical, Communication and Computer Engineering (ICECCE)\\
    30-31 October 2024, Kuala Lumpur, Malaysia
    \vspace{1.5em} 
    }}
\def\ps@IEEEtitlepagestyle{%
    \def\@oddhead{\rightconfheader} 
    \def\@evenhead{}%
    \def\@oddfoot{\footnotesize \rmfamily U.S. Government work not protected by U.S. copyright\hfill} 
}
\makeatother

\begin{document}

\bibliographystyle{IEEEtran}

\title{Handwriting Anomalies and Learning Disabilities through Recurrent Neural Networks and Geometric Pattern Analysis\\
}

\author{\IEEEauthorblockN{Vasileios Alevizos \orcid{0000-0002-3651-2134}}
\IEEEauthorblockA{\textit{Karolinska Institutet} \\
Solna, Sweden \\}
\and
\IEEEauthorblockN{Sabrina Edralin \orcid{0009-0003-6225-6351}}
\IEEEauthorblockA{\textit{University of Illinois Urbana-Champaign} \\
Illinois, USA \\}
\and
\IEEEauthorblockN{Akebu Simasiku}
\IEEEauthorblockA{\textit{Zambia University} \\
Ndola, Zambia \\}
\and
\IEEEauthorblockN{Dimitra Malliarou}
\IEEEauthorblockA{\textit{IntelliSolutions} \\
Athens, Greece \\}
\and
\IEEEauthorblockN{Antonis Messinis}
\IEEEauthorblockA{\textit{HEDNO SA} \\
Athens, Greece \\}
\and
\IEEEauthorblockN{George A. Papakostas \orcid{0000-0001-5545-1499}}
\IEEEauthorblockA{\textit{MLV Research Group, Department of Informatics, Democritus University of Thrace} \\
Thrace, Greece \\}
\and
\IEEEauthorblockN{Clark Xu \orcid{0000-0002-2384-4244}}
\IEEEauthorblockA{\textit{Mayo Clinic Artificial Intelligence \& Discovery} \\
Minnesota, USA \\}
\and
\IEEEauthorblockN{Zongliang Yue \orcid{0000-0001-8290-123X}}
\IEEEauthorblockA{\textit{Auburn University Harrison College of Pharmacy} \\
Alabama, USA \\}
}

\maketitle

\begin{abstract}
Dyslexia and dysgraphia are learning disabilities that profoundly impact reading, writing, and language processing capabilities. Dyslexia primarily affects reading, manifesting as difficulties in word recognition and phonological processing, where individuals struggle to connect sounds with their corresponding letters. Dysgraphia, on the other hand, affects writing skills, resulting in difficulties with letter formation, spacing, and alignment. The coexistence of dyslexia and dysgraphia complicates diagnosis, requiring a nuanced approach capable of adapting to these complexities while accurately identifying and differentiating between the disorders. This study utilizes advanced geometrical patterns and recurrent neural networks (RNN) to identify handwriting anomalies indicative of dyslexia and dysgraphia. Handwriting is first standardized, followed by feature extraction that focuses on baseline deviations, letter connectivity, stroke thickness, and other anomalies. These features are then fed into an RNN-based autoencoder to identify irregularities. Initial results demonstrate the ability of this RNN model to achieve state-of-art performance on combined dyslexia and dysgraphia detection, while showing the challenges associated with complex pattern adaptation of deep-learning to a diverse corpus of about 33,000 writing samples.
\end{abstract}


\section{Introduction}

\subsection{Poor Handwriting Association with Learning Disabilities}

Dyslexia and dysgraphia are two distinct learning disorders that significantly impact an individual's ability to read, write, and process language. Although each disorder presents its own challenges, they can sometimes coexist, compounding the difficulties. Both conditions exhibit unique visual patterns.

Dyslexia, predominantly a reading disorder, manifests itself through noticeable struggles with word recognition and spelling, paired with difficulties in phonological processing—the ability to link sounds with corresponding letters and words. Individuals with dyslexia frequently reverse letters and words \cite{helland_trends_2022}, causing them to mix up characters like b and d. It is also common for them to skip words or even entire lines while reading, disrupting comprehension. Words standing alone, stripped of contextual clues, often become unrecognizable, and reading becomes a slow and laborious task due to the additional mental effort required to decode each word.

Dysgraphia, rooted in the brain's wiring, hampers the ability to write coherently. Letter formation, spacing, and alignment of text become persistent issues that individuals often struggle with from an early age, despite dedicated practice and interventions \cite{gargot_acquisition_2020}. This disorder also affects fine motor skills, further complicating handwriting. Dysgraphia typically manifests as poor letter formation, with letters that are either misshapen or illegible. Inconsistent spacing appears both within words and between them, causing the text to become jumbled and disorganized. People with dysgraphia frequently mix uppercase and lowercase letters or vary the size of their characters significantly.

The co-occurrence of these conditions compounds the difficulties in reading, writing, and processing language, making day-to-day tasks challenging. A child or adult grappling with both dyslexia and dysgraphia may reverse letters while reading and then reproduce those same mistakes in their writing, further compounding the confusion. Tasks such as comprehending written instructions, producing legible assignments, or even communicating clearly through written words become an uphill battle.

\subsection{Visual Diagnosis of Dyslexia and Dysgraphia}

Visual pattern recognition of dyslexia and of dysgraphia may appear straightforward to the human observer, but automating detection with machine learning is much more challenging. Writing styles can vary significantly among individuals, influenced by factors such as mood and concentration, making analysis difficult. Grammatical errors also serve as crucial indicators, playing a significant role in the diagnostic process. Since different languages have their own distinctive scripts and letter formations, there is no universal solution that can apply to every detection case, which further increases the required solution granularity. As modern educational trends increasingly shift from handwriting to digital methods, diagnosing these conditions in digital texts may become even more challenging. Assessing dyslexia, for instance, requires a holistic approach led by qualified professionals like psychologists, educational diagnosticians, or speech-language pathologists. These experts conduct thorough evaluations that may encompass handwriting assessments, standardized testing, and detailed interviews. In terms of handwriting alone, individuals with dyslexia often reveal irregular letter formations, reversals, inconsistent sizing, and poor spacing between words and letters.

Meanwhile, dysgraphia creates a whole other set of distinct challenges when it comes to writing. As a learning disability that disrupts one's ability to produce coherent and readable handwriting, dysgraphia brings an idiographic set of patterns that can include irregular letter sizes, inconsistent spacing, and awkwardly slanted characters. Dysgraphia causes individuals to frequently cross out words or excessively erase as they attempt to write, and their grip on the pencil is often too tight, impairing their control over the writing movement. The advancement of technology brings a rising trend that continues to evolve; there is a growing trend toward minimizing the emphasis on natural handwriting in favor of digital tools. While this shift opens up new opportunities for people with learning disorders to express themselves, it could make detecting dyslexia and dysgraphia through traditional writing samples much trickier. CNN-based models show promising results \cite{yogarajah_deep_nodate}. Despite the progress made in automated diagnosis, there is room for future experimentation. In another study, authors rely on handwritten and geometric features through a Kekre-Discrete Cosine Transform model, attaining an impressive 99.75\% accuracy \cite{devi_dysgraphia_2023}. Also, incorporating diverse data modalities \cite{kunhoth_automated_2024} could improve accuracy. Quality features may generalize the model \cite{rashid_dysign_2023}. Further analytical features could capture sociocultural factors \cite{lomurno_deep_2023}, \cite{danna_tools_2023}. However, while the performance is generally high, they heavily rely on monotonic data sources such as focusing machine detection on highly standardized single-letter and evaluating detection models on these limited datasets of a few hundred writing samples.

\subsection{Contributions of Artificial Systems in Detection}

\begin{figure}[!htb]
    \centering
    \includegraphics[width=0.6\linewidth, keepaspectratio]{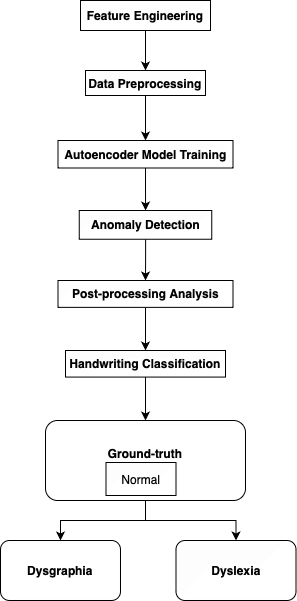}
    \caption{Workflow of handwriting anomalies detection starting with an inspection of relevant features that will later be used in the upcoming steps of anomaly detection.}
    \label{fig:workflow}
\end{figure}

With artificial intelligence promising to improve visual diagnosis, emerging applications have enabled the rapid scanning of multiple samples to pinpoint patterns that indicate potential learning disabilities \cite{richard_dyslexia_2020}. Researchers have harnessed deep learning (DL) and machine learning (ML) methodologies to enhance detection accuracy and support earlier identification. One study focusing on dysgraphia \cite{agarwal_early_2023} employed feature extraction of critical features tested on the Random Forest model; they achieved remarkable accuracy in detecting signs of this writing disability with limited availability of training data. Another study \cite{alkhurayyif_deep_2023} facilitated dyslexia detection with image processing and feature extraction, and the selected model, MobileNet V2, coupled with a single-shot detection (SSD) system, resulted in impressive precision. Further research proposed \cite{mikyska_handwriting_2023} a system that could assist those with dysgraphia to correct handwriting issues with a combination of neural networks and spelling and grammar correction algorithms. Further research \cite{t_automated_2024} explored pattern recognition techniques to reveal the subtle variations indicative of dysgraphia. As technology has rapidly developed, an increasing number of resources have become available to help individuals with dyslexia and dysgraphia including recording devices, portable notetakers, scanning and reading pens, text-to-speech systems, CD-based talking dictionaries, concept mapping software, as well as word prediction and word banks \cite{draffan2007use}, in which oftentimes individuals utilize a combination of these tools and some may not be readily available or known to individuals with dyslexia and dysgraphia. From an educator standpoint, there are primarily three overarching forms of assistive technologies (AT) employed to help students: 1) operating systems (or system preferences), 2) apps, and 3) extensions for web browsers \cite{dawson2019assistive}. In this study, we experiment to identify handwriting anomalies associated with dyslexia and dysgraphia to extend prior single-condition detection work to multi-condition detection and expand the impact of assistive technologies by creating early, automated detection over a robust and diverse set of writing samples. Via ML interpretability analysis, the workflow (\textbf{Fig. \ref{fig:workflow}}) involves a detailed inspection of writing features that play a significant role in the subsequent steps of classification.

Our approach began with image preprocessing to standardize the images and minimize biases using various techniques. This included resizing, converting to grayscale, and normalizing the images. The standardized images were then subjected to feature extraction. More specifically, baseline deviation (where characters deviate from a standardized writing line), connectivity issues (when letters are improperly spaced), and stroke thickness variations were identified as indicative features. Features were then inspected for shapes, letter angles, and overall writing density. Letter rotation was assessed for more than 45 degrees or if there was notable inconsistency in stroke thickness. The extracted features were compiled into a dictionary that captured critical information like stroke width variability and word shape variability, highlighting instances of handwriting impairments. The core classification process involved a neural network model tailored to analyze the extracted features and categorize the handwriting. Particularly, a custom RNN-based autoencoder structure was used to analyze the handwriting data. The encoder compressed high-dimensional inputs into a lower-dimensional feature representation while the decoder reconstructed the data to its original form. This autoencoder learned patterns and irregularities specific to dysgraphia and dyslexia. Finally, the model output, combined with the extracted features, was processed through a loss function that assessed handwriting against predefined criteria. It considers not only conventional classification loss but also additional penalties for severe baseline deviation and letter connectivity issues. In essence, modeling handwriting anomalies using diverse features enables the identification and differentiation of various pattern types.

\section{METHODOLOGY}
\subsection{Dataset}

In this study, we investigated handwriting anomalies across different conditions using a comprehensive dataset. The datasets \cite{ramlan_potential_2023}, \cite{drotar_dysgraphia_2020}, \cite{sihwi_dysgraphia_2019}, \cite{patricia_a_flanagan_nist_2016}, \cite{rosli_development_2021}, \cite{seman_notice_2021}, \cite{isa_cnn_2021}, \cite{isa_automated_2019} comprised over 33,000 images, ranging from single characters to full sentences, and was divided into two sets, 80\% train set, and 20\% test set, respectively. Our samples represented a range of handwriting patterns, including normal, dyslexia, low potential dysgraphia, and potential dysgraphia.

\subsection{Building Blocks and Ingredients}

\begin{figure}[!htb]
    \centering
    \includegraphics[width=\linewidth, height=0.8\textheight, keepaspectratio]{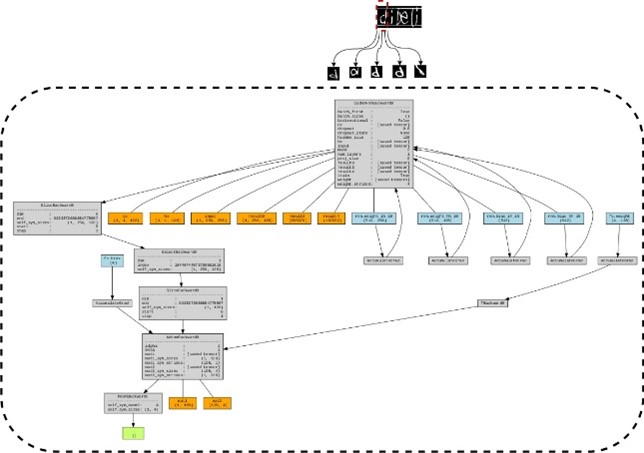}
    \caption{Computational graph visualization of a Recurrent Neural Network during back propagation.}
    \label{fig:rnn_backprop}
\end{figure}

The architecture, an RNN (Recurrent Neural Network) classifier \cite{schuster_bidirectional_1997} with an LSTM (Long Short-Term Memory) \cite{hochreiter_long_1997} layer, was selected due to its ability to capture sequential data. LSTM networks, with their internal memory mechanism, are particularly well-suited for processing handwriting patterns, like letter formation, baseline deviations, and other temporal features. To bolster its efficacy, we also incorporated an autoencoder structure (\textbf{Fig. \ref{fig:rnn_backprop}}). Autoencoders compressed and then reconstructed data, enabling the model to learn nuanced differences in writing patterns. The encoder uses convolutional layers to reduce input dimensions while retaining essential features, and the decoder reverses this process to restore features for classification. Each convolutional layer's output is fed to the subsequent LSTM layer, capturing both spatial and temporal features. For preprocessing, images were converted to grayscale to eliminate unnecessary color information. The dataset was divided into two subsets, ensuring that model performance can be gauged accurately. Each image was resized to a consistent 256x256 resolution. During training, adaptive moment estimation (Adam optimization) \cite{zhang_improved_2018} iteratively updates model weights, refining predictions by minimizing loss.

\subsection{Process of Text Irregularities}

The training process involves feeding batches of standardized handwriting images into the model, which extracts features and learns patterns associated with different handwriting conditions. Each batch is reshaped and fed through the RNN layers, which output classifications that guide the learning process. The confusion matrix, accuracy, and precision are computed after every training epoch to track the model’s performance. Confusion matrices are particularly useful in this context as they illustrate how well the model differentiates between dyslexia, dysgraphia, and normal handwriting.

\subsection{Evaluation}
For evaluation, we measured the accuracy and precision of classification,  identifying the true positive rate among the predicted positives. The F1 score combines precision and recall into one measure within a confusion matrix.

\section{Results}
Our data revealed that the average mean intensity for normal handwriting was approximately 0.19, with a histogram sum averaging around 12,671. Despite the standard deviation of mean intensity being relatively high, the variation in intensity showed significant differences between individual samples. The baseline deviation analysis for this class showed consistent handwriting flow and a notable vertical transition pattern with low variance.

In analyzing dyslexia-affected handwriting, we observed some distinct trends that set this group apart. The mean intensity in this group was lower than the normal group, at 0.17, with the histogram sum also reflecting a reduced value. A closer look into the standard deviation showed less consistent character formation and structure, which corroborates the common difficulties associated with recognizing letter patterns. These traits can be pivotal in refining feature extraction techniques. The vertical transitions for dyslexic handwriting were also markedly different, with erratic variations in writing lines due to frequent word skipping and letter reversals. In handwriting associated with low potential dysgraphia, we noted a significant reduction in mean intensity, reaching only 0.078 on average, with the histogram sum peaking around 5,100. The standard deviation of intensity was noticeably lower than other categories, indicating less variation in writing patterns. The characters' vertical transitions demonstrated a reasonably consistent formation. However, the uniformity was disrupted by occasional spikes in texture variance and baseline deviations.

Potential dysgraphia handwriting exhibited a slightly higher mean intensity than the low potential category, with the average hovering at around 0.107. The histogram sum was higher too, reaching an average of approximately 6,989. The vertical transitions remained more consistent than low potential dysgraphia but still indicated a general irregularity in the writing lines. This condition's baseline deviation and poor letter formation stood out, making the classification easier. From this analysis, several trends emerged that shape the feature extraction process. For one, the pronounced variations in mean intensity and histogram sum between the four classes form critical features to distinguish the different handwriting types. For example, reduced mean intensity was consistently associated with impaired writing abilities. The unique differences in vertical transitions, baseline deviations, and texture variance underscore the need for careful examination of spatial relationships between letters, which were often skewed in dysgraphia and dyslexia cases.

\begin{figure}[!htb]
    \centering
    \includegraphics[width=\linewidth, height=0.8\textheight, keepaspectratio]{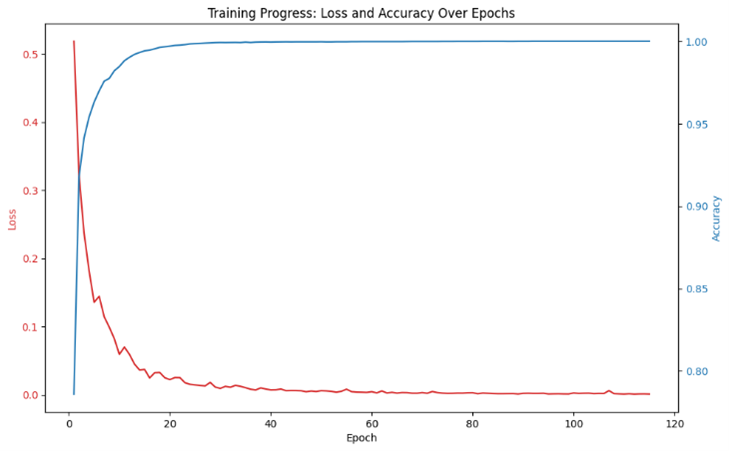}
    \caption{Analysis of training dynamics showing loss decrease and accuracy increase over 120 epochs, rapid improvement is present at the beginning of training followed by stabilization as accuracy sharply increases.}
    \label{fig:training_dynamics}
\end{figure}

\begin{figure}[!htb]
    \centering
    \includegraphics[width=\linewidth, height=0.8\textheight, keepaspectratio]{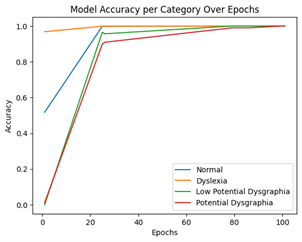}
    \caption{Comparative analysis of model accuracy across different handwriting conditions over epochs. Initially, accuracy increases across all categories. Dyslexia and ground truth adapted fastest; low potential dysgraphia adapted slowest; normal handwriting performed best overall.}
    \label{fig:comparative_analysis}
\end{figure}

The model's proficiency in differentiating between the typical handwriting and the other two main cases of dysgraphia and dyslexia progressively enhances over training epochs, which underscores the importance of both the quality of the training data and generalization on the presented features (as seen in \textbf{Fig. \ref{fig:training_dynamics}}). Initially, there was an increase of accuracy and a corresponding reduction in loss during the early epochs. However, as training progresses, a plateau in performance enhancement is observed. Once the model had assimilated the primary features, it encountered difficulties in further refining its predictions based on subtler differences in the data, which were less represented and more complex to learn. A slower adaptation rate corresponds to no significant improvement for the minority classes. High accuracy was towards the normal class (\textbf{Fig. \ref{fig:comparative_analysis}}), as evidenced by the confusion matrix, where very few normal cases were misclassified as dyslexia or dysgraphia. However, while there was a steady enhancement in the specificity for dyslexia and dysgraphia, the recall for these classes did not improve markedly. The observed differential adaptation speeds could be attributed to several factors. Primarily, the more robust or distinct feature representations of the normal class likely facilitated quicker learning compared to the subtler features of dyslexia and dysgraphia, which required more complex patterns within the network for effective learning. Furthermore, the imbalance in class sizes significantly impacts learning dynamics, as the model might overfit the majority normal class while neglecting the minority dyslexia and dysgraphia classes. In conclution, our pipeline, ameliorates disparate handwriting challenges, offering a ~ 1\% improvement over CNN-based approaches. The intractable nature of sequential data is better captured by our architecture's abstruse ability to handle temporal dependencies.

\section{DISCUSSION AND FUTURE DIRECTIONS}
Over time, RNNs have shown good performance in sequence prediction tasks (as seen in \textbf{Fig. \ref{fig:comparative_analysis}}) Comparative analysis of model accuracy across different handwriting conditions over epochs showed an initial increase in accuracy across all categories. Dyslexia and ground truth adapted fastest, low potential dysgraphia adapted slowest, and normal handwriting performed best overall, but struggle with long-term dependencies and scaling due to the vanishing gradients problem. Adding Convolutional Neural Networks (CNNs) or self-attention mechanisms, like those present in Transformers, can significantly improve performance by enhancing feature extraction and adaptability across diverse classes.

Future research could explore the integration of additional data modalities and the application of more complex neural architectures to enhance adaptability. Additionally, further investigation into the emotional aspects of writing could reveal whether changes in writing style across different emotions help identify specific patterns. Integrating advanced feature extraction techniques reveals new characteristics about pauses between strokes or variations in writing pressure, as both differ across individuals with dyslexia or dysgraphia.

\section{CONCLUSION}

Complex hand-writing styles associated with dyslexia and dysgraphia could be eased encomporating RNNs with autoencoders. By focusing on critical handwriting features such as baseline deviations, letter connectivity, and stroke thickness, the model effectively captures the nuances of these learning disabilities. CNNs, while highly effective in image-based analysis, tend to focus on spatial feature extraction and are limited in handling temporal dependencies that are essential for handwriting analysis. Relying only on isolated character recognition, simplifies the problem but misses the broader context of handwriting flow and consistency. By contrast, RNNs, tend to capture these sequential relationships better by adapting homogeneous patterns.




\vspace{12pt}

\end{document}